\documentclass{webofc}
\usepackage{listings}
\usepackage{hyperref}
\usepackage{graphicx}
\usepackage[utf8]{inputenc}
\usepackage{tikz}

\usepackage[draft]{minted}
\newminted{cpp}{gobble=2, numbersep=3pt, escapeinside=||, xleftmargin=5mm, framesep=2mm, frame=leftline, baselinestretch=1, fontsize=\scriptsize, linenos=true}
\newmintinline{cpp}{}
\makeatletter
\AtBeginEnvironment{minted}{\dontdofcolorbox}
\def\dontdofcolorbox{\renewcommand\fcolorbox[4][]{##4}}
\makeatother

\usepackage[varg]{txfonts}

\begin{document}

\title{ROOT I/O compression improvements for HEP analysis}
        \author{\firstname{Oksana}
        \lastname{Shadura}\inst{1}\fnsep\thanks
        {\email{oksana.shadura@cern.ch}}
        \firstname{Brian Paul}
        \lastname{Bockelman}\inst{2}\fnsep\thanks
        {\email{bbockelman@morgridge.com}}
        \firstname{Philippe}
        \lastname{Canal}\inst{3}\fnsep\thanks
        {\email{pcanal@fnal.gov}}
        \firstname{Danilo}
        \lastname{Piparo}\inst{4}\fnsep\thanks
        {\email{dpiparo@cern.ch}} \and
        \firstname{Zhe}
        \lastname{Zhang}\inst{1}\fnsep\thanks
        {\email{zhan0915@huskers.unl.edu}}
}

\institute{University of Nebraska-Lincoln, 1400 R St, Lincoln, NE 68588, United States
\and
           Morgridge Institute for Research, 330 N Orchard St, Madison, WI 53715, United States
\and
           Fermilab, Kirk Road and Pine St, Batavia, IL 60510, United States
\and
           CERN, Meyrin 1211, Geneve, Switzerland
}

\abstract{%
We overview recent changes in the ROOT I/O system, increasing performance and enhancing it and improving its interaction with other data analysis ecosystems. Both the newly introduced compression algorithms, the much faster bulk I/O data path, and a few additional techniques have the potential to significantly to improve experiment's software performance.

The need for efficient lossless data compression has grown significantly as the amount of HEP data collected, transmitted, and stored has dramatically increased during the LHC era. While compression reduces storage space and, potentially, I/O bandwidth usage, it should not be applied blindly: there are significant trade-offs between the increased CPU cost for reading and writing files and the reduce storage space.
}
\maketitle
\section{Introduction}
\label{intro}

In the past years LHC experiments are commissioned and now manages about an exabyte of storage for analysis purposes, approximately half of which is used for archival purposes, and half is used for traditional disk storage. Meanwhile for HL-LHC storage requirements per year are expected to be increased by factor 10 \cite{rnd}.

Looking at these predictions, we would like to state that storage will remain one of the major cost drivers and at the same time the bottlenecks for HEP computing. It means that new storage and data management techniques, as well as compression algorithms, are likely to be required to remove a cost bottleneck together with storage and analysis computing costs to be able to handle expected data ratios and data volumes needed to be processed by experiments during HL-LHC\cite{rnd}.

Looking into innovative compression algorithms could help to resolve some problems, such as improving user analysis, removing decompression speed bottleneck, while maintaining the same or better compression data ratios. Zstandard \cite{zstd} is a dictionary-type algorithm (LZ77) with a large search window and fast implementations of entropy coding stage, using either fast Finite State Entropy (tANS) or Huffman coding. Zstandard referred to as zstd, is a much more modern compressor comparing to Zlib, which was initially implemented in 1995, and which offers higher compression rates while using less CPU compared to other compression algorithms. ZSTD is available as a ROOT supported compression algorithm, starting from ROOT 6.20.00 release. \cite{root}

\section{Background}
\label{background}

Three years ago, Facebook \cite{facebook} open-sourced Zstandard, an innovative data compression solution that offers a  performance. It is largely supported by the community and continuously supported as well as enhanced by ZSTD authors, who released a variety of advanced capabilities, such as improved decompression speed and better compression ratios.

The initial promise of Zstandard was that it would allow users to replace their existing data compression implementation, such as ZLIB, for one with significant improvements on compression speed, compression ratio, and as well decompression speed. \cite{zstdandart} 

In addition to replacing ZLIB, ZSTD has taken over many of the tasks that traditionally relied on fast compression alternatives. Fastest compression is still provided by LZ4 (for the fastest compression settings), while ZSTD provides a twice size better compression ratio. According to reports from the community, it is slowly replacing the strong compression scenarios previously served by XZ (or LZMA) \cite{lzma}, with the benefit of 10 times faster decompression speed. According to reports from Facebook, with all these use cases combined, ZSTD now is processing a significant amount of data every day at Facebook.

Zstandard can use a "dictionary" format to make compression of files of an already known type in a more efficient way. Here a dictionary is a file that stores the compression settings for small files. Compression dictionary is assembled from a group of typically small files that contain similar information, preferably over 100 files. For the best efficiency, their combined size should be about one hundred times the size of the dictionary produced from them. In general, the smaller the file, the greater the improvement in compression. According to the zstd manual page, a dictionary can only increase the compression of a 64KB file by 10 percent, compared with a 500 percent improvement for a file of less than 1KB \cite{zstdandart}.

\section{Evaluation of simple ZSTD algorithm for LHC datatsets}
\label{evaluation}

\begin{figure}[!h]
  \centering
  \includegraphics[width=\textwidth]{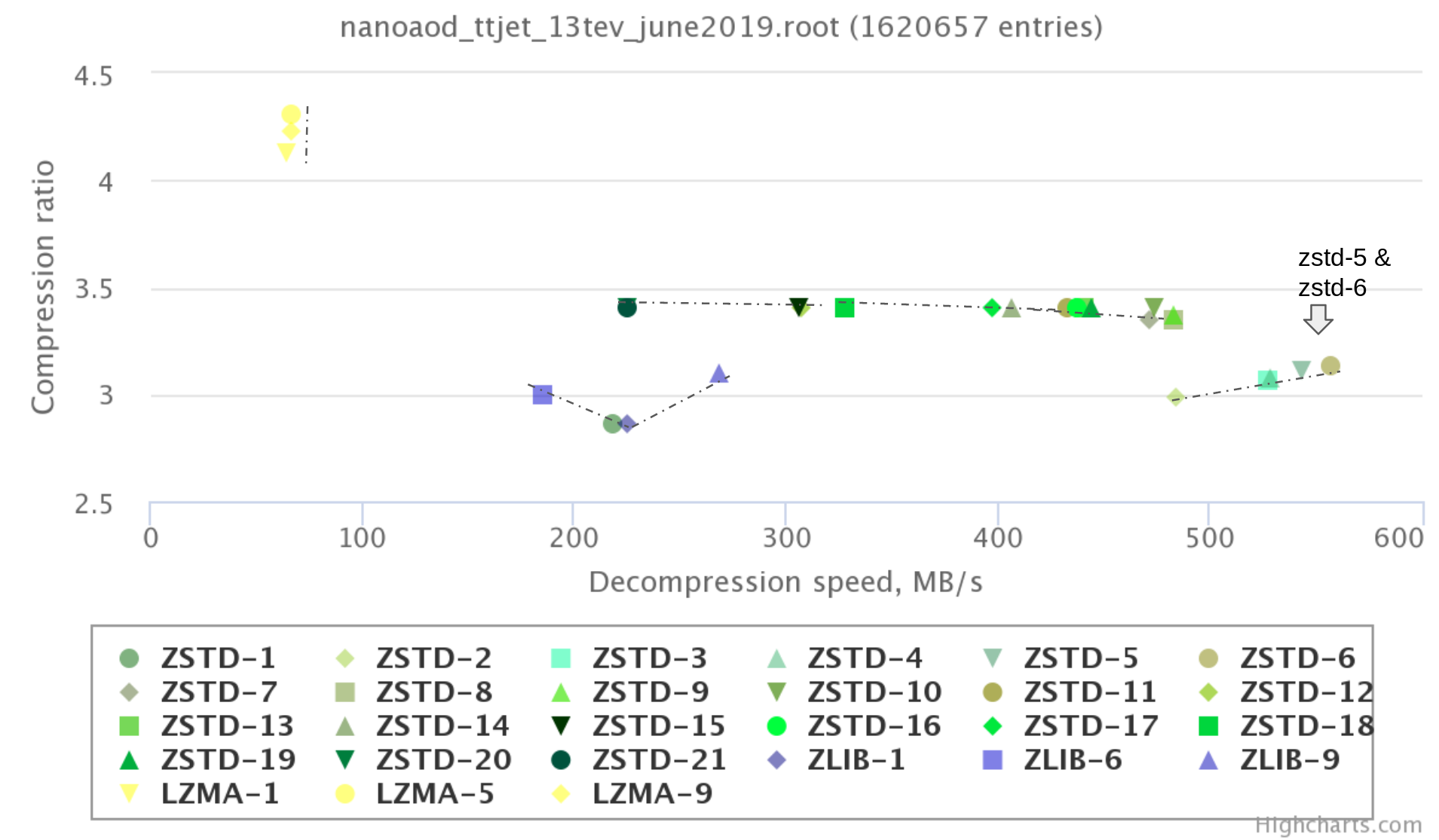}
  \caption{Comparison of compression ratio and decompression speed for ZLIB, LZMA and ZSTD algorithms for NanoAOD 2019 file}
  \label{figgenio}
\end{figure}

In this section, we will try to focus on the evaluation of compression of most used analysis-related formats in CMS, NanoAOD \cite{nanoaod} and MiniAOD \cite{miniaod}, as well as a simple case of analysis file used by the LHCb experiment.

The MiniAOD is a new high-level CMS data file that was introduced in 2014 to serve the needs of the mainstream physics analyses while keeping a small event size - only 30-50 KB per event. It is not readable with bare ROOT and requires special CMSSW setup to be able to read it. Meanwhile, NanoAOD format consists of a Ntuple like format, readable with bare ROOT and containing the per-event information that is needed in most generic analyses. The size per event of NanoAOD is the order of 1KB. NanoAODs are usually centrally produced or even produced on-demand with different variations of features or columns required by different physics analysis groups. Users can as well easily extend NanoAOD for their specific studies making a private production when needed.

\begin{figure}[!h]
  \centering
  \includegraphics[width=\textwidth]{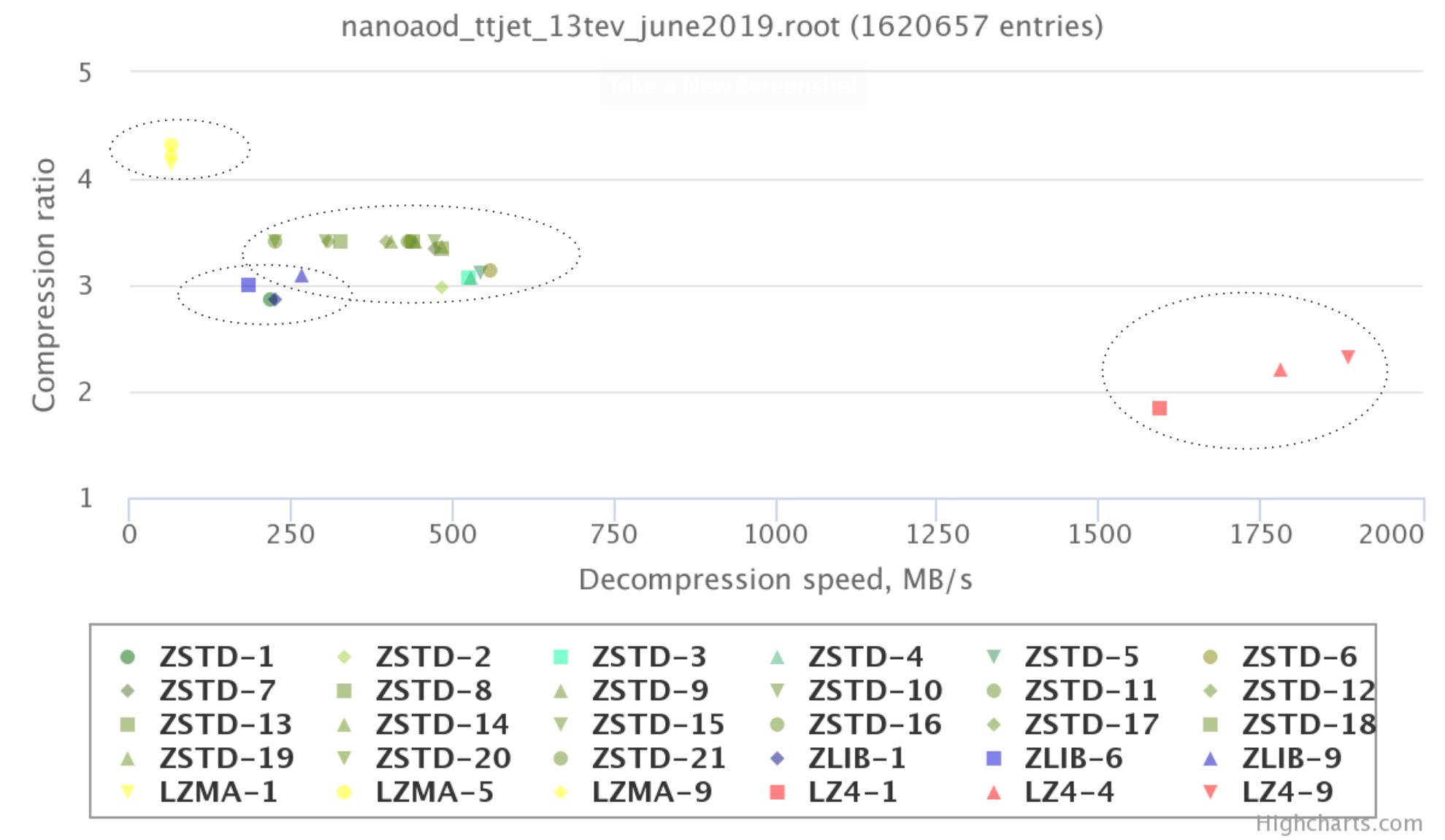}
  \caption{Comparison of compression ratio and decompression speed for all compression algorithms for NanoAOD 2019 file.}
  \label{figionanoad}
\end{figure}

For CMS NanoAOD files, using ZSTD could be a better compromise between size of file on a disk and decompression speed for a faster analysis as well as better compression ratio and \textit{2x} faster decompression then ZLIB and 6x faster comparing to LZMA, while file compressed with ZSTD is only 20 \% bigger size (all results are shown on the Figure \ref{figionanoad} and \ref{figgenio}).

For  MiniAOD, measured time spend in decompressing on readback is \textit{15x} less comparing to LZMA, while the size of the file with ZSTD is only 10\% bigger.

\begin{figure}[!h]
  \centering
  \includegraphics[width=\textwidth]{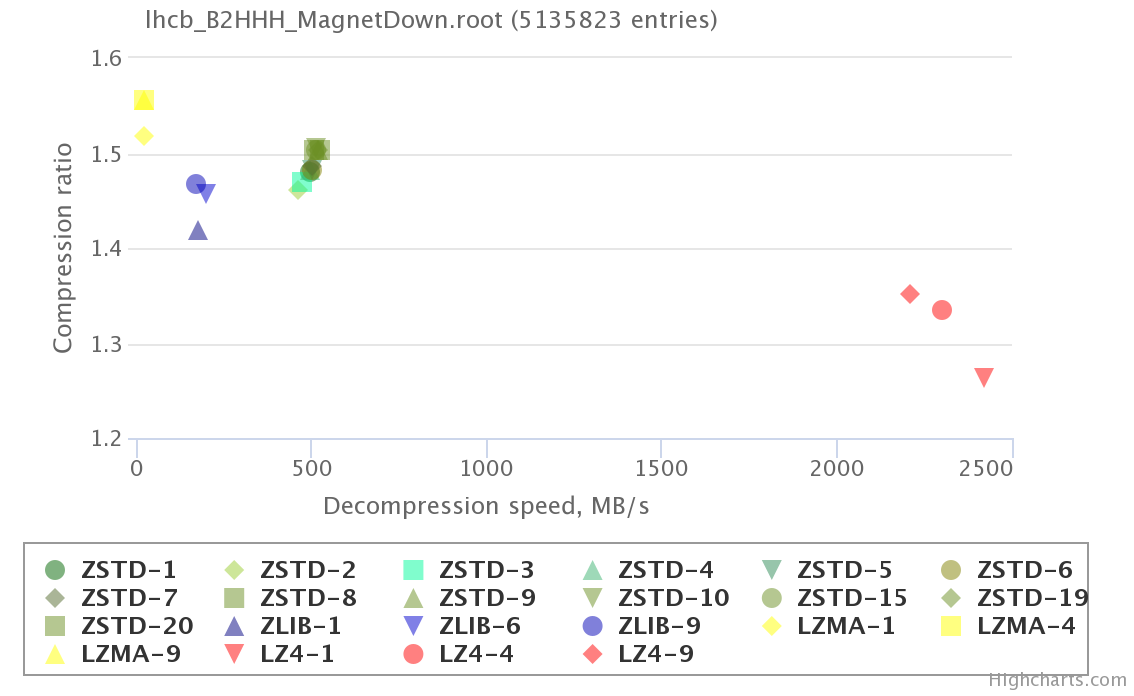}
  \caption{Comparison of compression ratio and decompression speed for all compression algorithms for LHCb file.}
  \label{figiolhcb}
\end{figure}

In case of LHCb, for the very simple NTuples with a simple structure, the best choice could be LZ4 compression algorithm, offering \textit{10x} time faster read speed (all results are shown on the Figure \ref{figiolhcb}).

\begin{figure}[!h]
  \centering
  \includegraphics[width=\textwidth]{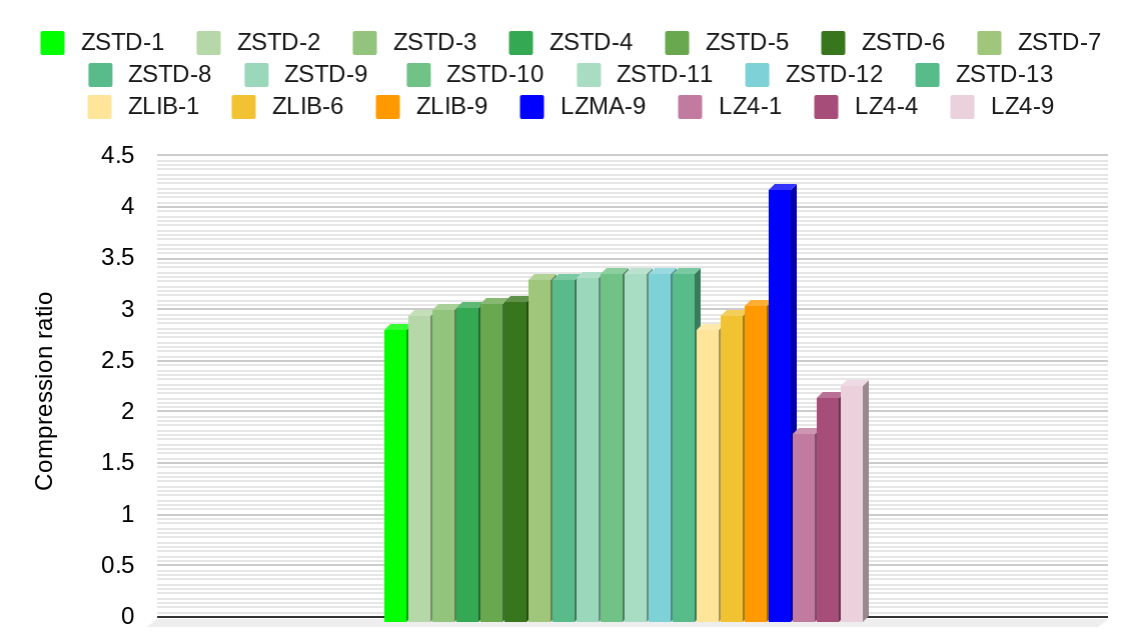}
  \caption{Comparison ratio comparison for custom analysis file with variable-sized data (containing C-style arrays).}
  \label{figiocstyle}
\end{figure}

In ROOT, the serialization of variable-sized data (containing C-style arrays) produces two internal arrays: one contains the branch data for each of events while the other contains the byte offset of each of events in the branch data. LZ4 compression algorithm achieves its performance by looking for byte-aligned patterns (as opposed to ZLIB compression algorithm, which works on individual bits) and lacks the Huffman encoding pass, this results in the offset array sequence being effectively incompressible using LZ4. ZSTD has no problems with compression of data that contains the byte offset of each event in the branch data (vs LZ4) (all results are shown on the Figure \ref{figiocstyle}).

\section{TTree::kOnlyFlushAtCluster option, offering faster decompression}

\begin{figure}[!h]
  \centering
  \includegraphics[width=0.8\textwidth]{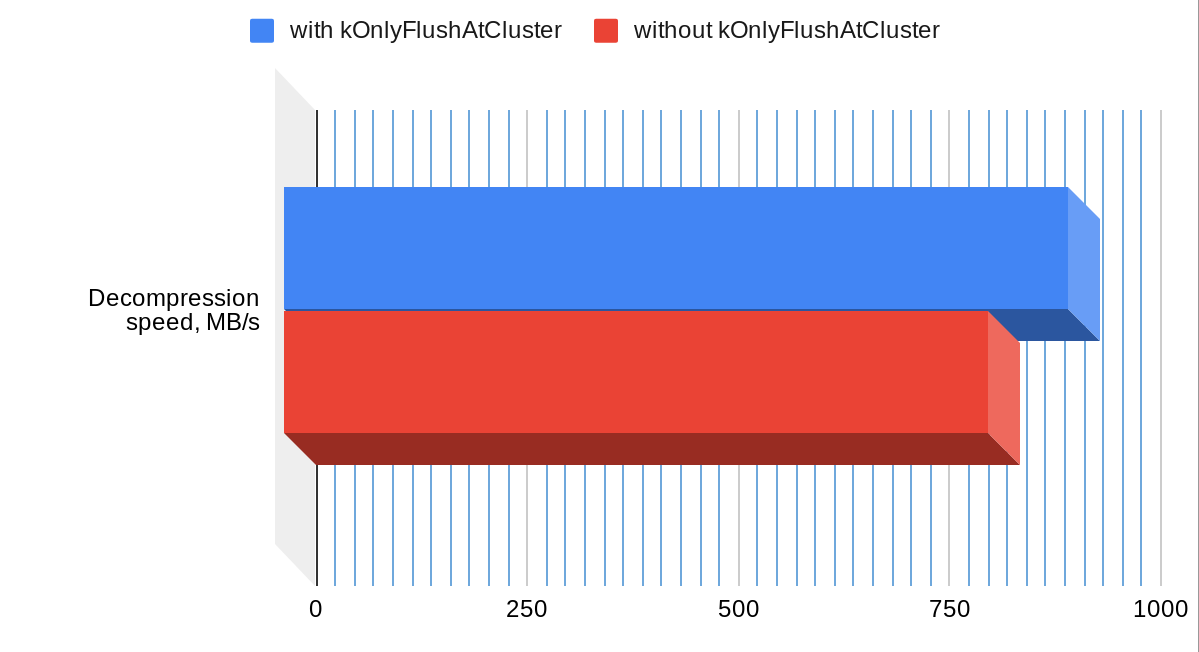}
  \caption{Comparison of decompression speed for two file samples NanoAOD 2017, with and without $TTree::kOnlyFlushAtCluster$ option.}
  \label{figioflush}
\end{figure}

TTrees can be forced to create only the new baskets at event cluster boundaries, using a \textit{TTree::kOnlyFlushAtCluster} feature. It simplifies file layout and I/O at the cost of memory. For example for the \textit{TTree::kOnlyFlushAtCluster} feature tests shown in Figure \ref{figioflush}, NanoAOD 2017 was bigger only by 3.6 \% of size, while decompression speed is improved almost up to 200 MB/s \cite{rootio}. 

$TTree::kOnlyFlushAtCluster$ is recommended for simple file formats such as ntuples where it can show really interesting improvementsts, but not for more complex data types.

\section{Limitations and Future work}
\label{limitationsandfuture}

Some time ago, Bitshuffle pre-conditioner was demonstrated as a possible pre-conditioner for ROOT data with LZ4 for lossless compression. To improve the performance of LZ4 in this case, we investigated the combination of LZ4 with various “pre-conditioners”. Pre-conditioners transform the sequence of input bytes according to a simple, deterministic algorithm before applying the compression algorithm. The two algorithms investigated, inspired by the Blosc library, are Shuffle and BitShuffle. Both pre-conditioners rearrange the input array’s bytes by reading through the data using fixed strides. The resulting output of the pre-conditioner often contains long sequences of repeated bytes, improving the compression ratio for LZ4. One of the issues exposed was that it is difficult for ROOT to compress its buffers now due to its 9-byte header \cite{rootio}.

The idea of using pre-conditioners could be easily expanded to be used with other algorithms, such as ZSTD. The next goal of the project will be to validate the possibility to use pre-conditioners in the ROOT compression layer used to compress both ROOT file formats (TTree and RNTuple) for the fastest ROOT compression algorithms: LZ4, ZSTD.

Another interesting investigation could be to extend pre-conditioners to support new BYTE\_STREAM\_SPLIT encoding that improves compression ratio and compression speed for certain types of floating-point data where the upper-most bytes of values do not change much. The existing compressors and encodings in ROOT do not perform well for such data due to noise in the mantissa bytes. The new encoding improves results by extracting the well compressible bytes into separate byte streams which can be afterward compressed by a compressor like ZSTD \cite{parquet}.

\section{Conclusions}
\label{conclusion}

ZSTD has been successfully evaluated and ready to be used for compression of data analysis formats for future LHC Runs in experiments.

\section{Acknowledgments}
\label{ackn}
This work has been supported by U.S. National Science Foundation grants OAC-1450323.

\end{document}